\documentclass[amsmath,amssymb,aps,prd,nofootinbib,twocolumn]{revtex4-2}

\usepackage{graphicx}
\usepackage[utf8]{inputenc}
\usepackage{dcolumn}
\usepackage{bm}
\usepackage{xcolor}
\usepackage[colorlinks = true, linkcolor = purple, urlcolor  = blue, citecolor = blue, anchorcolor = blue]{hyperref}
\usepackage{float}
\usepackage[caption=false]{subfig}
\usepackage{multirow}
\usepackage{array}
\newcolumntype{P}[1]{>{\centering\arraybackslash}p{#1}}

\def\barnue{\mathrel{{\bar \nu}_e}}

\def\t13{\mathrel{{\theta_{13}}}}
\def\y12{\mathrel{{\tan^2 \theta_{12}}}}
\def\c2{\mathrel{{\chi^2 }}}

\newcommand{\gw}{GW}
\newcommand{\n}{neutrino}
\newcommand{\ns}{neutrinos}
\newcommand{\sn}{supernova}
\newcommand{\sne}{supernovae}
\newcommand{\mms}{multi-messenger}
\newcommand{\mm}{memory}

\newcommand{\gws}{GWs}
\newcommand{\df}{DSNB}
\newcommand{\nsn}{NSFC}
\newcommand{\fsn}{BHFC}

\newcommand{\be}{\begin{equation}}
\newcommand{\ee}{\end{equation}}
\newcommand{\ba}{\begin{eqnarray}}
\newcommand{\ea}{\end{eqnarray}}

\begin{document}

\title{Memory-triggered supernova neutrino detection}

\author{Mainak Mukhopadhyay}
\email{mmukhop2@asu.edu}
 \affiliation{ Department of Physics, Arizona State University
PO Box 871504, Tempe, AZ 85287-1504, USA.}
\author{Zidu Lin}%
 \email{zlin23@utk.edu}
\affiliation{Department of Physics and Astronomy,
University of Tennessee 
Knoxville, TN 37996-1200, USA}
\author{Cecilia Lunardini}%
 \email{Cecilia.Lunardini@asu.edu}
\affiliation{%
 Department of Physics, Arizona State University
PO Box 871504, Tempe, AZ 85287-1504, USA
}%

\date{\today}

\begin{abstract}
We demonstrate that observations of the gravitational memory from core collapse supernovae at future Deci-Hz interferometers enable time-triggered searches of supernova neutrinos at Mt-scale detectors. Achieving a sensitivity to characteristic strains of at least $\sim 10^{-25}$ at $f\simeq 0.3$ Hz -- e.g., by improving the noise of DECIGO by one order of magnitude -- will allow robust time triggers for supernovae at distances $D\sim 40-300$ Mpc, resulting in a nearly background-free sample of $\sim 3-70$ neutrino events per Mt per decade of operation. This sample would bridge the sensitivity gap between rare galactic supernova bursts and the cosmological diffuse supernova neutrino background, allowing detailed studies of the neutrino emission of supernovae in the local Universe. 
\end{abstract}

\maketitle

\section{Introduction}
Neutrinos are major players in the emerging field of \mms\ astronomy. With gravitational waves (\gws) and photons, they have the potential to probe the most extreme astrophysical phenomena in unprecedented detail. Core collapse supernovae (CCSNe) are prime targets of \mms\ observations, where neutrinos dominate the energy output and carry direct information on the extremely dense environment surrounding the collapsed core.
The $\sim 10$ s burst of \ns\ from a \sn\ will also allow tests of particle physics beyond the Standard Model~\cite{Raffelt:1987yt,Turner:1987by,Mayle:1987as,Chang:2018rso}.
\begin{figure}
      \includegraphics[width=0.48\textwidth,angle=0]{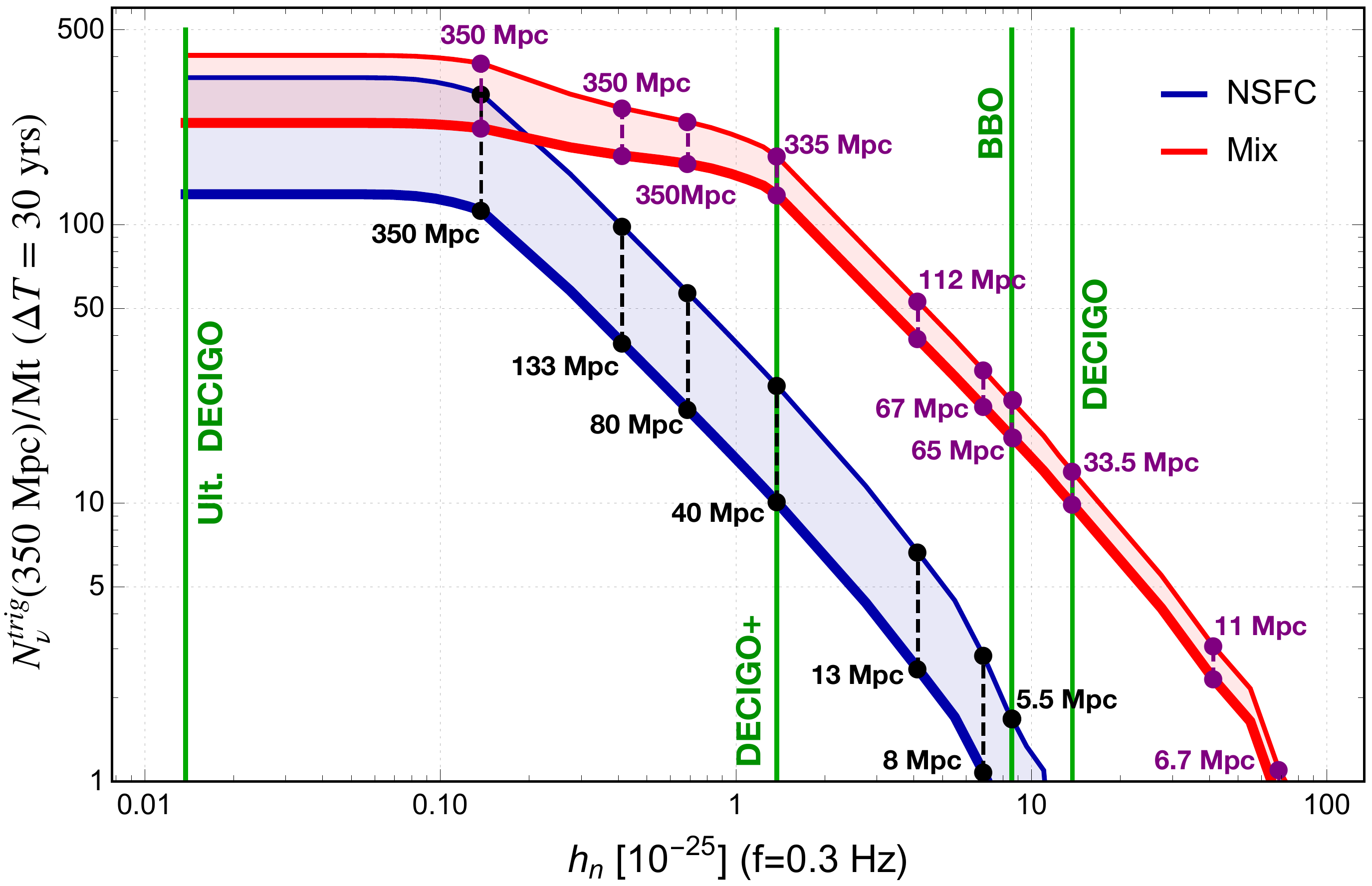}
  \caption{The number of memory-triggered \sn\ \ns\ detected at a 1 Mt water Cherenkov detector in 30 years, as a function of the noise of the GW detector at $f=0.3$ Hz. 
  The vertical lines mark specific experiments considered here. 
  The lower and upper shaded regions refer respectively to a homogeneous population with moderate memory strain and a mixed population where 40\% of collapses have stronger memory strain; 
  the shading describes the effect of varying the neutrino spectrum, see Table \ref{tab:parameters}. The dots (upper set: NSFC and lower set: BHFC) and legends on the curves give the GW distance of sensitivity ($r^{GW}_{max}$, see text below Eq.~\eqref{eq:snr}) corresponding to the noise on their abscissa.
 }
\label{fig:nnutot_d}
\end{figure}

\begin{figure}
      \includegraphics[width=0.45\textwidth,angle=0]{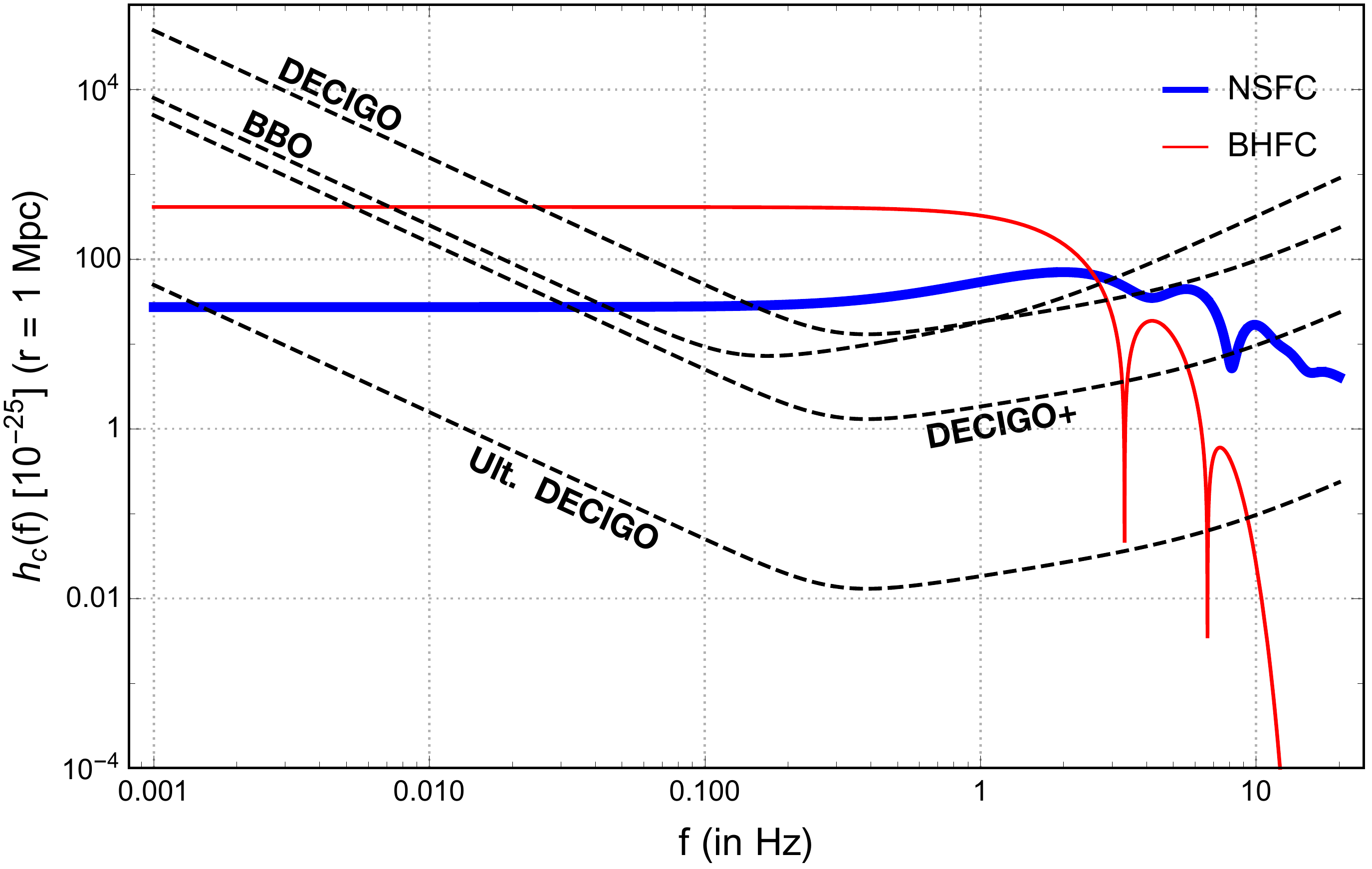}
  \caption{Solid: the characteristic gravitational memory strain $h_c(f)$ for the \nsn\ and \fsn\ models (thin and thick lines respectively). The distance to the \sn\ is $r=1$ Mpc. Dashed: sky-averaged noise curves for representative detectors (see  fig. \ref{fig:nnutot_d}).
 }
\label{fig:hcf_models}
\end{figure}

The detection of an individual \sn\ \n\ burst is exciting as well as challenging. A statistically significant observation is possible only for \sne\  within 1-3 Mpc from Earth \citep{PhysRevLett.95.171101,Kistler:2008us},
where collapses are rare, resulting in decades of waiting time. An alternative is to search for the Diffuse Supernova Neutrino Background (DSNB), from all the \sne\ in the universe \citep{1982SvA....26..132B,Krauss:1983zn,Beacom:2010kk,2017hsn..book.1637L}, which has a substantial cosmological component.  ${\mathcal O}(10-100)$ DSNB \ns\  could be detected in a decade 
(see, e.g., \cite{DeGouvea:2020ang}), and 
preliminary data could be available in just a few years \citep{Beacom:2003nk,Super-Kamiokande:2013ufi,JUNO:2015zny,Hyper-Kamiokande:2018ofw,Theia:2019non,abi2020deep}. 

Burst and DSNB searches lack sensitivity to the local universe, $r\sim 3 -100 $ Mpc, where many supernova-rich galaxies are situated. Ideas to overcome this gap typically rely on time-triggers that would allow to identify a single neutrino as signal instead of background. One could use either a neutrino self-trigger — where $2-3$ neutrinos observed less than 10 s apart can be attributed to a supernova with high confidence \citep{PhysRevLett.95.171101,Adams:2013ana} — , or the time coincidence with the $O(10^2)$ Hz supernova \gw\ signal from interferometers like LIGO-Virgo and its successors \citep{PhysRevLett.103.031102,LIGOScientific:2016jvu,Super-Kamiokande:2021dav}. Both methods are still limited to a few Mpc distance, except for the most optimistic GW models (see, e.g.,~\cite{szczepanczyk2021detecting} and references therein) and futuristic multi-Megaton \n\ detectors \citep{Kistler:2008us}\footnote{Astronomical observations of supernovae can not serve as time triggers, due to the ${\mathcal O}(1)$ hour uncertainty in the time delay between the neutrino and the electromagnetic signal from the same star.}.

\begin{figure*}[t!]
\begin{center}
      \includegraphics[width=0.47\textwidth,angle=0]{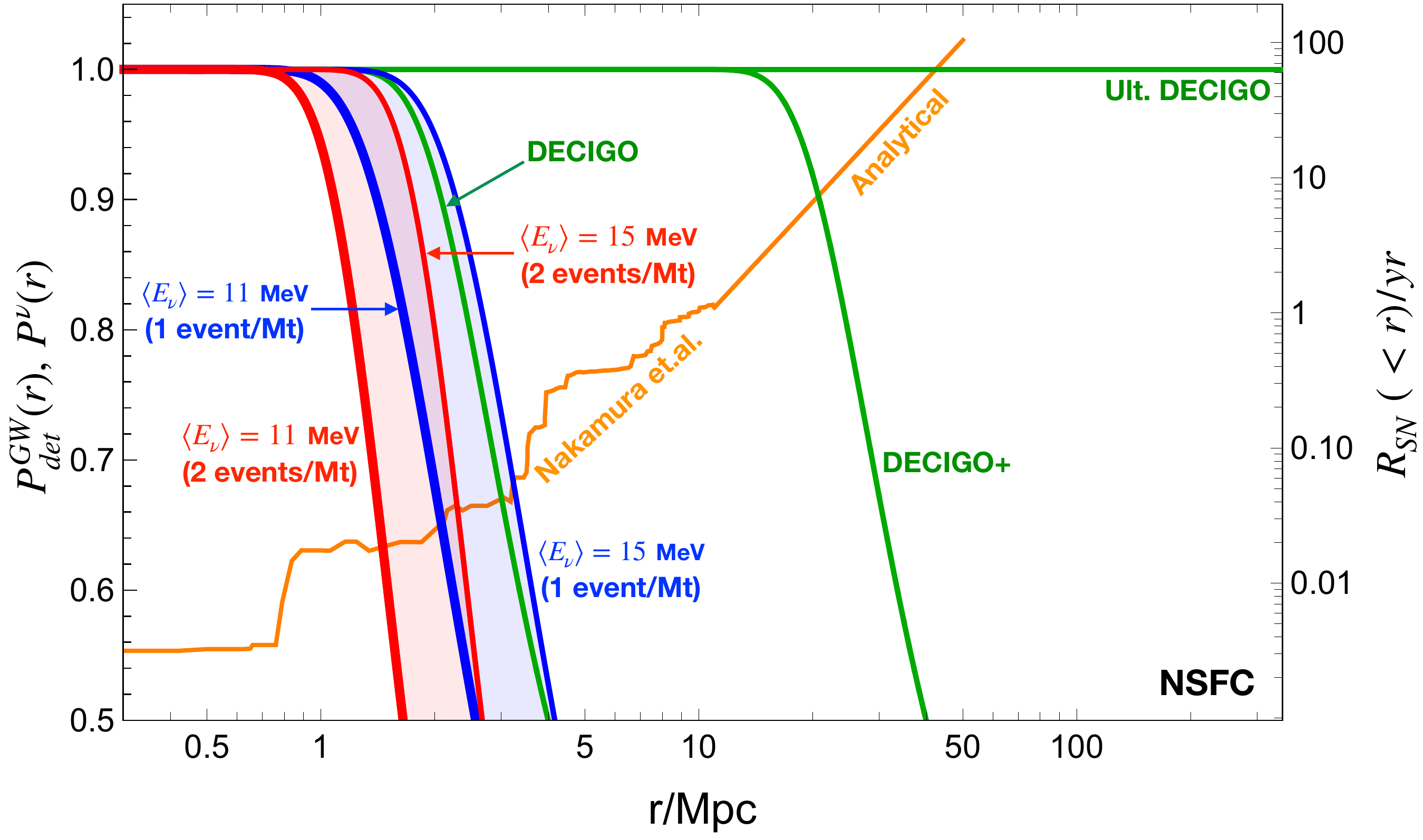}
      \includegraphics[width=0.47\textwidth,angle=0]{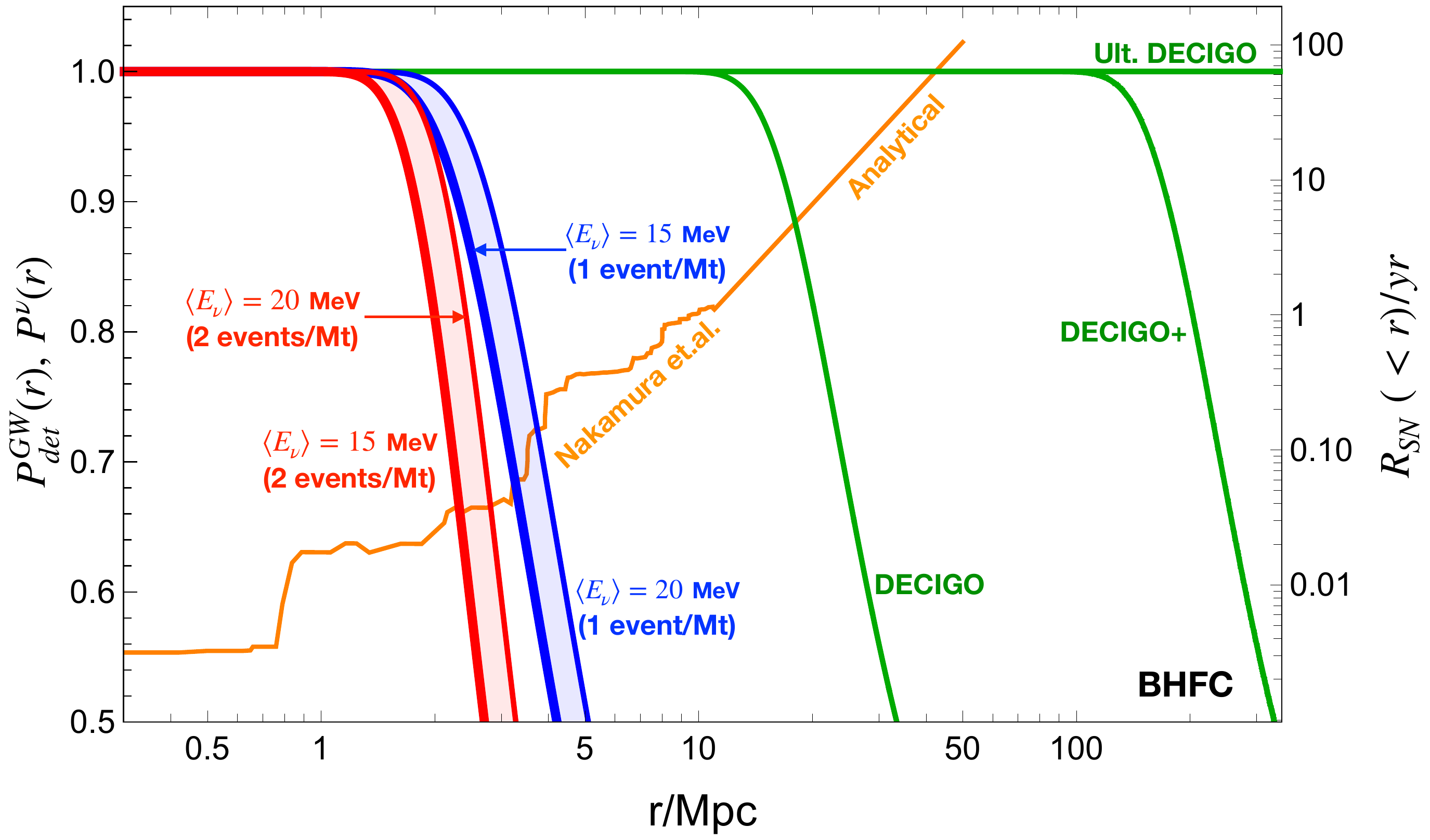}
  \caption{Detection probabilities for a memory signal, $P^{GW}_{det}(r)$, at three \gw\ detectors of reference, and neutrino detection probabilities, $P^\nu(1,r)$ and  $P^\nu(2,r)$ (see eq. (\ref{eq:pnu})). Shadings describe the variations with the varying \n\ spectrum, see Table \ref{tab:parameters}. The left (right) panel is for \nsn\ (\fsn).  Also shown is the cumulative rate of core collapses (vertical axis on the right). See labels on the curves for details. 
  }. 
\label{fig:rsn_pdet}
\end{center}
\end{figure*}

In this paper, we propose a new time-triggered method to detect supernova \ns, which is potentially sensitive to \sne\ up to $\sim$ 100 Mpc. 
The time trigger is the observation of the gravitational memory signal caused by the neutrino emission itself. The memory is a non-oscillatory, permanent distortion of the local space time due to the anisotropic emission of matter or energy by a distant source. The \mm\ due to \n\ emission by a \sn\ at distance $r$ has characteristic strain $h_c \sim 10^{-23}-10^{-21} (\mathrm{10~kpc}/r)$ and frequencies in the Deci-Hz band, $f\sim 0.1 - 3$ Hz \citep{Burrows:1995bb,1997A&A...317..140M,kotake2009ray,Muller:2011yi,Li:2017mfz,Vartanyan:2020nmt,Richardson:2021lib}. The memory develops  $\sim 0.1$ s from the start of the \n\ emission, thus being an ideal time-trigger. Next generation powerful Deci-Hz \gw\ detectors, like the Deci-hertz Interferometer Gravitational wave Observatory (DECIGO) \citep{Seto:2001qf,PhysRevD.83.044011,Sato:2017dkf,2021PTEP.2021eA105K} and the Big Bang Observer (BBO) \citep{PhysRevD.83.044011} will provide robust triggers for \sne\ at 10 Mpc and beyond \citep{Mukhopadhyay:2021zbt}. These would result in a nearly pure sample of $\sim 10 - 100$  supernova neutrino events from the local universe within a few decades; see our summary figure, fig.\ref{fig:nnutot_d}.
Here we illustrate our proposed methodology and its physics potential.
\section{Formalism}
%
\subsection{Gravitational memory signals}
The \sn\ \n\ memory strain can be expressed as \citep{Epstein:1978dv,Turner:1978jj,1997A&A...317..140M}
\be
\label{eqn:mainampeq}
h^{xx}_{TT} = h(r,t) = \frac{2G}{r c^4} \int_{-\infty}^{t-r/c} dt^{\prime} L_{\nu}(t^{\prime}) \alpha(t^{\prime})\,.
\ee
where $c$ is the speed of light, $t$ is the time post bounce and $G$ is the universal gravitational constant. $L_\nu$ is the all-flavors \n\ luminosity and $\alpha\sim {\mathcal{O}}(10^{-3} - 10^{-2})$ is the time-varying anisotropy parameter~\cite{kotake2009ray,Vartanyan:2020nmt}\footnote{
In axisymmetric simulations~\cite{1997A&A...317..140M,kotake2009ray}, only the $'+'$ strain may be extracted, and the observer is positioned such that the $'+'$ strain is maximized.}.
Simulations  show that $\alpha(t)$ becomes non-zero within a few ms post-collapse, during the accretion phase, and can change sign multiple times within the first second, as a result of the dynamics of the matter near the collapsed core.
The behavior of $\alpha(t)$ at $t>1$ s, during the cooling phase, is unknown. 
Following \cite{Mukhopadhyay:2021zbt}, we consider two phenomenological models  for the 
\mm: the first, characterized by a weaker and shorter anisotropic phase, is representative of a neutron-star-forming collapse (\nsn); the second has larger and prolonged anisotropy, and could represent a black-hole-forming collapse (\fsn). In both models, $\alpha=0$ for $t>1$ s. Maximum  values of $r h(r,t)\sim 26.5$ cm and  $r h(r,t) \sim 400$ cm are obtained for the two models respectively.  In fig. \ref{fig:hcf_models}, we show the memory characteristic strain~\cite{Li:2017mfz}, $ h_c(r,f) = 2f|\Tilde{h}(r,f)|$, where $\Tilde{h}(r,f)$ is the Fourier Transform of $h(r,t)$.  Also shown are the noise curves of  Deci-Hz detectors, which are given by the quantity $h_n (f) = \Upsilon \sqrt{5\ f\ S_n (f)}$~\cite{Li:2017mfz}, where $S_n(f)$ is the power spectral noise density \citep{Sathyaprakash:2009xs}. We choose  $\Upsilon = 1, 10^{-1}, 10^{-3}$; the first and last correspond to DECIGO and its optimal (futuristic) realization, Ultimate DECIGO \citep{Seto:2001qf,PhysRevD.83.044011,Sato:2017dkf}; the middle value represents an hypothetical intermediate case (DECIGO+ from here on). 

The detectability of a \mm\ signal is determined by the signal-to-noise (SNR) ratio of the detector\footnote{Here the comparison with published SNR curves has indicative character only; a signal-specific study of the detectability is ultimately needed, and is left for future work. }, which is defined as~\citep{Moore:2014lga}
\be
\label{eq:snr}
\rho^2(r) = \int_{-\infty}^{\infty} d (\text{log} f) \Bigg( \frac{h_c(r,f)}{h_n(f)} \Bigg)^2\,. 
\ee
We compute the probability of detecting a CCSN \mm, $P_{det}^{GW}$, for a fixed false alarm probability $P_{FA}^{GW}=0.1$. This requires producing Receiver Operating Curves (ROCs) in the plane $P_{det}^{GW}-P_{FA}^{GW}$, which we do following the formalism in~\cite{Jaranowski:1999pd} for $N=3$ degrees of freedom (here $N$ is set  equal to the number of Gaussian functions used to represent $\alpha(t)$, see \cite{Mukhopadhyay:2021zbt})\footnote{In Ref. \cite{Jaranowski:1999pd}, the formalism of $P_{det}$ and $P_{FA}$ are presented in the context of matched filter analysis. In the search for gravitational memory signals, we applied a filter studied in \cite{Mukhopadhyay:2021zbt}, which reasonably reproduce the results from numerical simulations.}. The result is that $P_{det}^{GW}$, at a fixed $P_{FA}^{GW}$, is an increasing function of $\rho(r)$, through which it depends on the distance, $r$. We define the \gw\ detector distance of sensitivity, $r^{GW}_{max}$ such that $P^{GW}_{det}(r^{GW}_{max})=0.5$. 
$P_{det}^{GW}(r)$ is shown in fig. \ref{fig:rsn_pdet} for our cases of reference. For DECIGO, and for  \nsn\ and \fsn\ respectively, we have $r^{GW}_{max}\simeq 4$ Mpc and $r^{GW}_{max}\simeq 33$ Mpc. We find $r^{GW}_{max}\simeq 40$ Mpc and $r^{GW}_{max}\simeq 335$ Mpc for DECIGO+; for Ultimate DECIGO, $r^{GW}_{max} > 350$ Mpc for both population models.

%
We note in passing that, in principle, the stochastic effect of the memory signals from cosmological supernovae contributes to the noise in a \gw\ detector, and therefore to $r^{GW}_{max}$.  For real-time searches of transient signals at a modern interferometer like LIGO, the noise spectral density is measured over sliding time windows of ${\mathcal O}(10^2)$ s width, leading to a fast identification of seconds-long transients \cite{KAGRA:2021bhs}. 
Due to the high supernova rate ($\dot \rho_{SN}\sim 10^{-4}~\mathrm{ yr^{-1}~Mpc^{-3}}$ locally, corresponding to $\sim 10^7$ core collapses per year in the visible universe)~\cite{Madau:1998dg,Ando:2004hc,Daigne:2004ga}, the individual cosmological memory signals combine to constitute a continuum, that is best described by an integral over the cosmic volume. Such integral represents the contribution of supernovae to the fraction of cosmic energy density in GW, $\Omega_{GW}$ (see, e.g., \cite{2001astro.ph..8028P,Buonanno:2004tp,Crocker:2017agi,Finkel:2021zgf} for the formalism). 

Following Ref. \cite{Buonanno:2004tp}, we have estimated the supernova memory contribution to $\Omega_{GW}$, and found that it affects the probability curves  in fig. \ref{fig:rsn_pdet} solely for Ultimate DECIGO, and only for $r\gtrsim 300$ Mpc and for the most optimistic \mm\ model (BHFC curve in fig. \ref{fig:hcf_models}, corresponding to a GW spectral energy density $\Omega_{\mathrm GW}=\mathcal{O}(10^{-17})$). As will be seen in the next section, the triggered \n\ search is limited to $r\lesssim 300$ Mpc by the background at the \n\ detector. Therefore, the stochastic GW noise from supernovae is negligible in the present context, and will not be considered further.  



\subsection{Neutrino signals}

For \n\ detection, we consider  a water Cherenkov experiment, where the main channel of sensitivity is inverse beta decay (IBD), $\barnue+p \rightarrow n + e^+$.  For the time-integrated (over $\Delta t= 10$ s) $\barnue$ flux at Earth, $\Phi(r, E_\nu)$  
we use analytical quasi-thermal spectra of the form given in \cite{Keil:2002in}.  The average $\barnue$ energy is varied in an interval motivated by numerical simulations \citep{Sukhbold:2015wba,Ertl:2015rga,Kresse:2020nto}, in a way to effectively account for \n\ oscillations.  The spectrum shape parameter, $\beta$, and the total energy in $\barnue$ are fixed. See Table \ref{tab:parameters} for details.  

\begin{table}[H]
\begin{center}
\begin{tabular}{|| P{1.3cm} || P{1.2cm} | P{1.0cm} || P{1.7cm} || P{1cm} | P{1cm} ||}
\hline
\multicolumn{1}{||P{1.3cm} ||}{\textbf{Model}} & 
\multicolumn{2}{P{2.2cm}||}{\centering \textbf{Energy \\$(\times 10^{53} \text{ergs})$}} & 
\multicolumn{1}{P{1.7cm}||}{\centering $\beta$} & 
\multicolumn{2}{P{2.0cm}||}{\centering \textbf{$\langle E_\nu \rangle$ \\ (in MeV)}}\\
\hline
& Ac. ph. & $\bar{\nu}_e$ &  & Lower & Upper \\
\cline{2-3}
\cline{5-6}
\nsn\ & $1.2$ & $0.5$ & $3$ & $11$ & $15$ \\
\hline
\fsn\ & $2$ & $0.45$ & $2$ & $15$ & $20$ \\
\hline
\end{tabular}
\caption{The neutrino flux parameters, from numerical simulations \cite{Sukhbold:2015wba,Ertl:2015rga,Kresse:2020nto}. The Ac. ph. and $\barnue$ columns refer to the all flavor energy in the accretion phase only (which contributes to the \mm\ signal, see text) and to the energy in $\barnue$ emitted over the time-triggered interval of 10 s.  $\beta$ is related to the second momentum of the spectrum: $\beta=(2 \langle E_\nu \rangle^2 - \langle E_\nu^2 \rangle)/(\langle E_\nu^2 \rangle - \langle E_\nu \rangle^2)$. }
\label{tab:parameters}
\end{center}
\end{table}

The predicted number of  events in the detector from a CCSN at distance $r$ is: 
\be
N (r) = \int_{E^{th}_\nu}^{E^{max}_\nu} N_{p}  \eta  \sigma(E_\nu)  \Phi(r, E_\nu)\ dE_\nu \,,
\label{eq:nev}
\ee
where $N_{p}$ is the number of target protons, $\eta=0.9$ is the detection  efficiency~\citep{PhysRevD.38.448, Abe:2011ts, PhysRevD.97.103001} and $\sigma (E_\nu)$ is the IBD cross-section \citep{Strumia:2003zx}. We take an energy interval $\left[ E^{th}_\nu, E^{max}_\nu\right]=[19.3,50]~\mathrm{ MeV}$ to avoid the spallation background at low energy and the atmospheric \n\ background at high energy \citep{Abe:2011ts,Hyper-Kamiokande:2018ofw,phdthesis1}. 
We find $N(1~\mathrm{ Mpc})\simeq 5 - 12$ and $N(1~\mathrm{ Mpc})\simeq 12 - 18$  for \nsn\ and \fsn\ respectively, by varying the mean $\barnue$ energy in the intervals  given in Table~ \ref{tab:parameters}. 

The Poisson probability of observing $N \geq N_{min}$ neutrino events in a detector is
\be
P^{\nu}(N_{min},r) = \sum_{n = N_{min}}^{\infty} \frac{N^n(r)}{n!} e^{-N(r)}\,.
\label{eq:pnu}
\ee
It is plotted for $N_{min} = 1,~2$ in fig. \ref{fig:rsn_pdet} for the two models of reference. As expected, $P^{\nu}(N_{min},r)$ declines rapidly at $r \gtrsim 3$ Mpc. 
%

\begin{figure*}[t]
\begin{center}
    \includegraphics[width=0.48\textwidth]{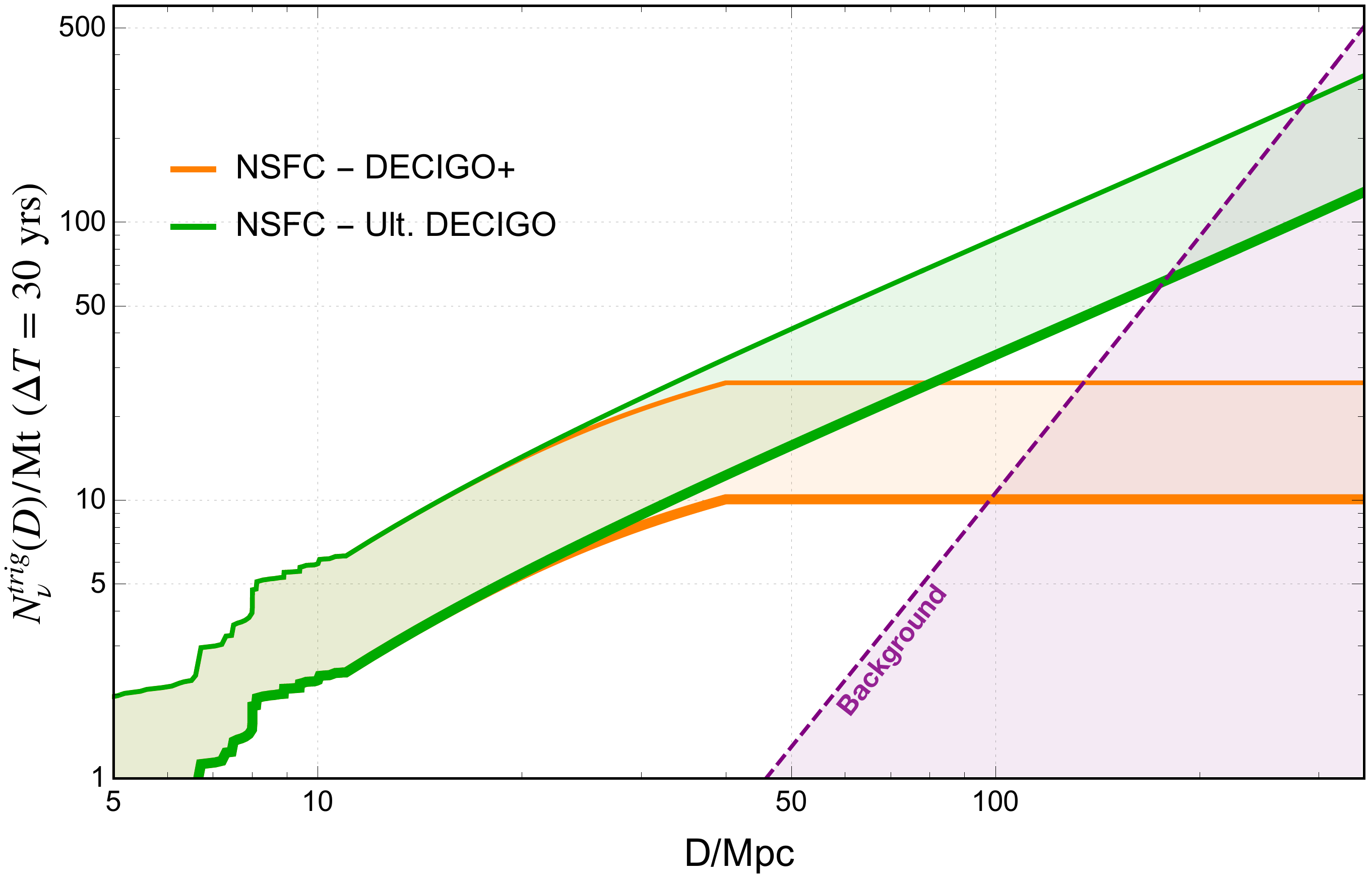} \hfill
    \includegraphics[width=0.48\textwidth]{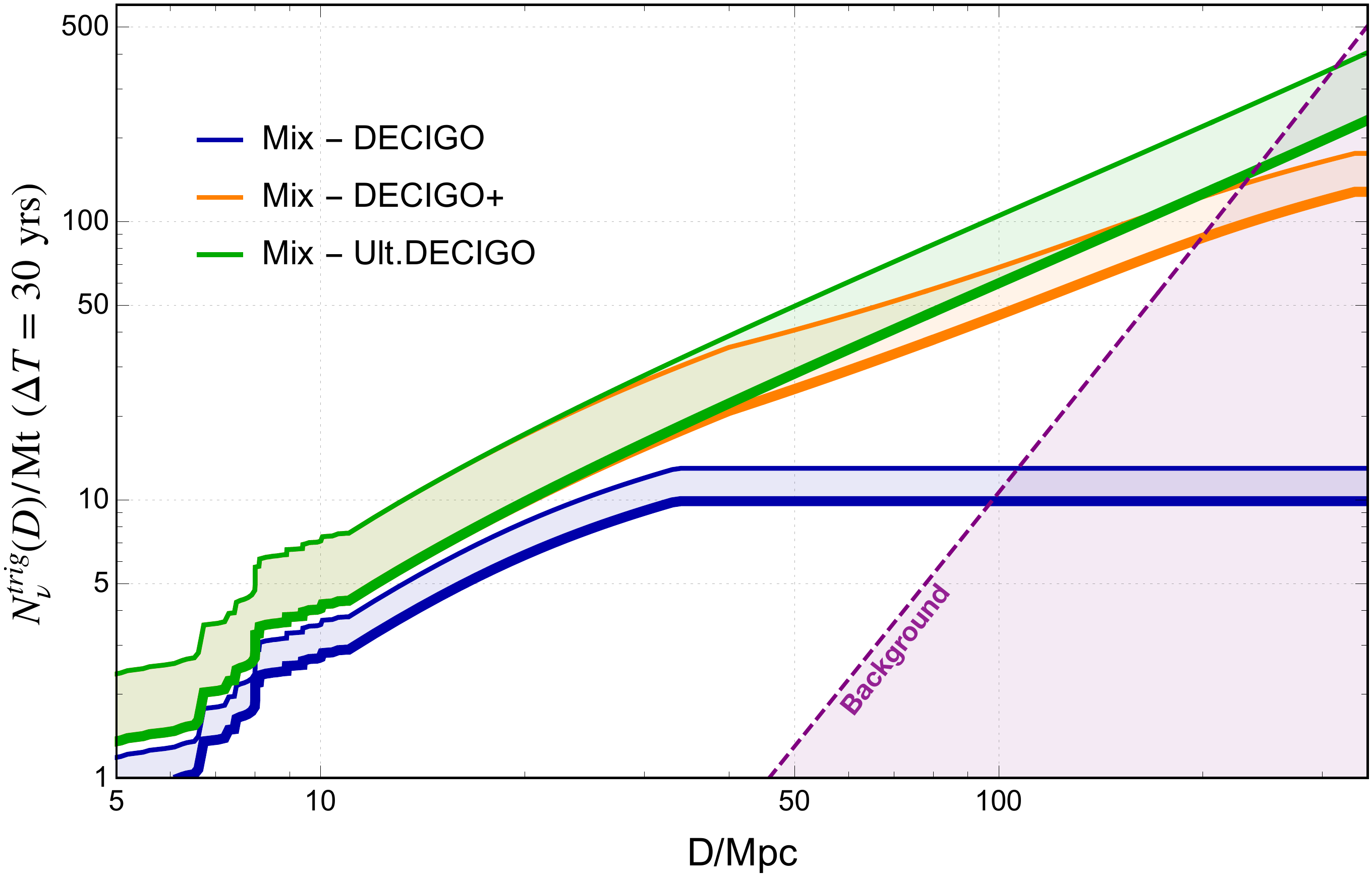} \hfill
    \caption{
    Number of background events and of memory-triggered \n\ events from collapses at distance $r<D$, as a function of $D$, for a  Mt water Cherenkov detector and 30 years running time. The upper to lower shaded regions are for triggers from Ultimate DECIGO, DECIGO+ and DECIGO (the latter is invisible in the left panel).  Shadings describe the effect of varying the \n\ spectrum, see Table \ref{tab:parameters}. Left panel: homogeneous \nsn\ population. Right: mix of 60\% \nsn\ and 40\% \fsn.  
    }
    \label{fig:nnu_d}
\end{center}
\end{figure*}



\section{Memory-triggered neutrino observations }

\subsection{Event rates}

To estimate the 
rate of  memory-triggered \n\ events, we model the rate of core collapses as a function of $r$.
For $r\lesssim 11$ Mpc, we use the rates for individual galaxies from~\cite{Nakamura:2016kkl}.
For $r> 11$ Mpc we assume a constant volumetric rate of $R_{SN}=1.5~10^{-4}~\mathrm{ Mpc^{-3} yr^{-1}}$ (the evolution with redshift is negligible for the distances of interest here).
%
The cumulative rate (total rate of core collapses with $r < D$) is shown in fig. \ref{fig:rsn_pdet}.

The number of \mm-triggered \n\ events from all \sne\ 
within a distance $D$, over a detector running time $\Delta T$ can be calculated as a sum over all the galaxies (index $j=1,2,...$), at distance $r_j<D$: 
\be
\label{eq:trigevents}
    N^{trig}_{\nu}(D)=\Delta T \sum_{j,r_j<D} R_j N(r_j)P_{det}^{GW}(r_j)\,,
\ee
where $R_j$ indicates the \sn\ rate in the galaxy $j$. This discrete expression is replaced by a continuum one, involving an integral, for $D>11$ Mpc, where the cosmological \sn\ rate is used.


We now discuss the background of the time-triggered neutrino search. The number of \sn\ \mm\ signals observed in the time $\Delta T$ is, $N^{trig}_{SN}(D)=\Delta T \sum_{j,r_j<D} R_j P_{det}^{GW}(r_j)$, and the number of expected background events is $N^{trig}_{bckg}(D)=N^{trig}_{SN}(D) \lambda \Delta t$, where $\lambda\simeq 1313$ events/year is the background rate in the detector \citep{Abe:2011ts,Hyper-Kamiokande:2018ofw,phdthesis1}. Note that the background level is reduced by a factor $\epsilon_{bckg=}N^{trig}_{SN}(D) \Delta t/\Delta T$ compared to an un-triggered search\footnote{The time delay effect due to the non-zero neutrino mass can be neglected; it is estimated to be only a fraction of a second for energies and distances of interest here, see, e.g., \cite{Burrows:1991kf}. }. 

We limit our study to  neutrino events (eq. (\ref{eq:trigevents})) from CCSNe in the cosmic volume with $4<D<350$ Mpc, thus accounting for the fact that a nearby \sn\ ($D<4$ Mpc) is unlikely to occur in three decades time. The upper bound on $D$ is justified  because beyond it the total event rate becomes dominated by background.
Experimentally, a distance cut can be accomplished in different ways. For \nsn, one can make a selection using estimates of $D$ from astronomy follow ups, which will benefit from the alerts from the \mm\ detection and should have excellent sensitivity to supernovae in the local universe (see, e.g., \cite{Kochanek:2017wud,Hiramatsu:2020obu,Valenti:2016uhk,spiro2014low,Hosseinzadeh:2017uig} for dedicated projects). In the absence of an optical counterpart (\fsn), a similar (although less efficient) data selection can be performed using minimal input from theoretical models, e.g. to obtain conservative upper limits on the distances of individual observed BHFCs via GW memory signals. In the mature stage of this search,  specifically designed data-analysis algorithms --  exploiting the correlation of multiple observables -- could reduce the level of model-dependency to a minimum. 
\subsection{Results}

Our main results are in fig. \ref{fig:nnu_d} and fig. \ref{fig:nnutot_d} for $\Delta T=30$ yrs and for two scenarios: (i) a \sn\ population entirely comprised of \nsn; and (ii) a mixed population with 60\% \nsn\ and 40\% \fsn.  
Fig. \ref{fig:nnu_d} shows $N^{trig}_{\nu}(D)$ as a function of $D$. We observe the (expected) trend $N^{trig}_{\nu}(D)\propto D$ for $D\lesssim r^{GW}_{max}$ \footnote{Recall that, in the continuum limit, the number of \sne\ scales like $D^3$ and the flux dilution factor like $D^{-2}$.}, 
with a flattening of the curves at larger $D$ due to the loss of sensitivity of the \gw\ detector. 
For case (i), time triggers from DECIGO+ will result in  $N^{trig}_{\nu}\sim 10-30$. For Ult. DECIGO, $N^{trig}_{\nu}\sim 100-300$ is expected\footnote{For comparison, our estimated number of CCSNe within 350 Mpc is $N^{trig}_{SN}\sim 1.21 \times 10^6$.}. 
For the mixed population (case (ii)), results for Ult. DECIGO change only minimally, due to the different \n\ parameters between \nsn\ and  \fsn. Instead, $N^{trig}_{\nu}$ increases dramatically, surpassing 100, for DECIGO+, due to the larger distance of sensitivity to \fsn. Indeed, the number of triggered \n\ events from collapses with $30 < D < 350$ Mpc is dominated by \fsn\ (see also fig.  \ref{fig:rsn_pdet}). For this mixed population scenario, even DECIGO could be effective, providing a few triggers of \fsn\ up to $D \sim 30$ Mpc, resulting in $N^{trig}_{\nu} \sim 10$. As fig. \ref{fig:nnu_d} shows, in all cases the signal exceeds the background for triggers with $r\lesssim 100$ Mpc. For Ultimate DECIGO, even for the largest $D$ the signal is comparable to the background, and would cause a statistically significant excess. 

Our summary figure, fig. \ref{fig:nnutot_d}, shows $N^{trig}_{\nu}(350~{\mathrm Mpc})$, as a function of $h_n$, together with representative values of $r^{GW}_{max}$. 
Roughly, we find $N^{trig}_{\nu} \propto 1/h_n$,
for $h_n \gtrsim 10^{-26}$, with a flattening at lower values of $h_n$, due to  upper cutoff on $D$. It appears that, even for the most conservative parameters, a ${\mathcal O}(10)$ noise abatement with respect to DECIGO (i.e., DECIGO+) is sufficient to obtain a signal at a Mt scale detector in $\sim 20-30$ years. 

\section{Conclusions and Discussion}

Summarizing, we have described a new multimessenger approach to core collapse \sne, where a time-triggered search of \sn\ \ns\ is enabled by observing the gravitational memory caused by the \ns\ themselves. This scenario could be realized a few decades from now, when powerful Deci-Hz interferometers (noise $h_n \lesssim 10^{-25}$) and Mt-scale neutrino detectors start operating. For optimistic parameters, DECIGO and HyperKamiokande (mass $M=0.260$ Mt) might already achieve a  low statistics observation. This approach will also enable  joint analyses of neutrino, GW and light curves of CCSNe in local universe.   
%



Our proposed method will deliver a sample of \n\ events from \sne\ in the \emph{local} universe, from which the main neutrino properties -- i.e, the (population-averaged) energy spectra and time profiles-- will be measured. These can then be compared to the same quantities from (1) SN 1987A, to measure the deviation between SN1987A and an average local supernova (the same exercise can be done for a future nearby supernova burst, if it occurs); (2) the \df, to distinguish the contributions to the \df\ by CCSNe in the distant universe and by other transients (e.g. binary mergers).
The comparison between cosmological and local contributions to the DSNB will test hypotheses of how the supernova progenitor population evolves with the distance. Even within the local-neutrino sample, one could test the evolution with distance, if the latter is estimated for each \sn\ using multi-messenger observations (e.g.,  the amplitude of the \mm\ signal and astronomical imaging).


Correlating \mm\ and \n\ data might reveal two distinct populations, like those described here (\nsn\ and \fsn), which could be statistically separated. For example, events having a relatively large neutrino-memory time separation (bigger than 1 s, as black hole formation typically occur within 1 s, cutting off the \n\ luminosity \citep{RevModPhys.74.1015,Sumiyoshi:2006id,OConnor:2010moj}) and(or) followed by electromagnetic (EM) signals of a CCSN could be attributed to NSFC. The possibility to study such sub-population individually is unique of this local-collapses \n\ sample. Additionally, our method provides a unique chance to jointly analyze neutrino and follow-up EM signals~\citep{2009arXiv0912.0201L,Kochanek:2017wud} from the \emph{same} NSFC. Although only $\approx1$ event would be detected from a specific NSFC, it can help to determine the time when the core of a NSFC collapses and the shock is formed. Such estimation would be relatively precise, 
considering that the \n\ burst from a NSFC only lasts for  $\approx10$ s.
A \sn\ EM signal is delayed  relative to the \ns, by at least the time
it takes the shock to propagate through the envelope, typically hours. 
Measuring this time delay will provide a crucial confirmation and can test the variation of the CCSNe explosion mechanism.

To conclude, we have demonstrated that the interplay between \n\ detectors and sub-Hz \gw\ observatories will open a new path to studying supernova neutrinos. 
%
Although several decades may pass before the first results become available, the work of designing the next generation of experiments is well under way, and we hope that our work will contribute to its progress.

\acknowledgements
We are grateful to Raffaella Margutti and Michele Zanolin for useful discussions and comments. We also thank the anonymous referees for suggesting valuable improvements. We acknowledge funding from the National Science Foundation grant number PHY-2012195. MM was also supported by the Fermi National Accelerator Laboratory (Fermilab) Award No. AWD00035045 during this work. ZL acknowledges funding from the NSF Grant No. PHY-1554876, PHY 21-16686 and from DOE Scidac Grant DE-SC0018232. 


\bibstyle{aps}
\bibliography{draft_mt}

\end{document}